\documentclass[12pt]{article}

\usepackage{amssymb}

\newcommand{\beq}[1]{\begin{equation}\label{#1}}
\newcommand{\eeq}{\end{equation}}
\newcommand{\bear}[1]{\begin{eqnarray}\label{#1}}
\newcommand{\ear}{\end{eqnarray}}
\newcommand{\nn}{\nonumber}

\catcode`\@=11 \@addtoreset{equation}{section}\catcode`\@=12

\newcommand{\R}{ {\mathbb R} }

\newcommand{\diag}{ \mbox{\rm diag} }
\newcommand{\e}{ \mbox{\rm e} }
\newcommand{\eps}{ \varepsilon }

\newcommand{\p}{\partial}

\newcommand{\tri}{\Delta}

\newcommand{\fnm}{\footnotemark}
\newcommand{\fnt}{\footnotetext}

\begin{document}

 \begin{center}
 \large \bf
  Quantum billiards in multidimensional models with fields of forms

 \end{center}

 \vspace{0.3truecm}

 \begin{center}

 \normalsize\bf

  V. D. Ivashchuk\fnm[1]\fnt[1]{e-mail: ivashchuk@mail.ru}
  and V. N. Melnikov\fnm[2]\fnt[2]{e-mail:  melnikov@phys.msu.ru},

 \vspace{0.3truecm}

 \it Center for Gravitation and Fundamental Metrology, VNIIMS, Ozyornaya St., 46, Moscow 119361, Russia and

 \it Institute of Gravitation and Cosmology, Peoples' Friendship
 University of Russia,  Miklukho-Maklaya St., 6, Moscow 117198,
 Russia

 \end{center}

\begin{abstract}

 Cosmological Bianchi-I type model in the $(n+1)$-dimensional
gravitational theory with several forms is considered. When
electric non-composite brane ansatz is adopted the Wheeler-DeWitt
(WDW) equation for the model, written in the conformally-covariant
form, is analyzed. Under certain restrictions asymptotic solutions
to WDW equation near the singularity are found which reduce the
problem to the so-called quantum billiard on the
$(n-1)$-dimensional Lobachevsky space $H^{n-1}$. Two examples of
quantum billiards are considered: $2$-dimensional quantum billiard
for $4$-dimensional  model with three $2$-forms and
$9$-dimensional quantum billiard for $D = 11$ model with $120$
$4$-forms which mimic $M2$-brane solutions of $D=11$ supergravity.
For certain asymptotic solutions the vanishing of the wave
function at the singularity is proved.

\end{abstract}


  \section{Introduction}

In this paper we deal with  the quantum billiard approach for
multidimensional cosmological-type  models defined on the manifold
 $(u_{-},u_{+}) \times \R^{n}$, where $n \geq 3$.

 In classical case the billiard approach was suggested by
 Chitre \cite{Chit} for explanation the
 BLK-oscillations \cite{BLK} in the  Bianchi-IX model
 \cite{Mis0,Mis1} by using a simple triangle billiard
 in the Lobachevsky space $H^2$.

 The billiard approach in dimension $D=4$ in classical and quantum case
 was also considered in papers  of A.A. Kirillov \cite{Kir-93}.

 In multidimensional case the
 billiard  representation for cosmological model with
 multicomponent  ``perfect'' fluid  was  introduced in
 \cite{IKM1,IKM2,IMb0}.  In   \cite{IMb0} the finiteness of
 the billiard volume  was formulated in terms of the so-called
 illumination problem.  The inequalities on Kasner parameters were
 written  and the quantum billiard was also considered.

 The  billiard approach for multidimensional models
 with scalar fields and fields of forms
 was suggested in  \cite{IMb1}, where
 inequalities on Kasner parameters were found. These inequalities
 played an important role in the proof of ``chaotic''   behavior
 in superstring-inspired   (e.g.  supergravitational)  models
 \cite{DamH1,DHN}.

 Recently the quantum billiard approach for a multidimensional
 gravitational model with several forms was considered in
 \cite{KKN,KN}. The asymptotic solutions to WDW equation presented
 in these papers are equivalent to those  obtained earlier in \cite{IMb0}.
 In refs. \cite{IMb1,KKN,KN} a semi-quantum approach was
 used, when gravity was quantum but the matter (e.g. fluids, forms)
 was considered as the classical one. It should be noted that
 a semi-quantum form of WDW equation for the model with fields of forms and
 scalar field was suggested earlier in \cite{LMMP}.

 Here we use another form of the WDW equation with enlarged
 minisuperspace which include the form potentials
 $\Phi^s$ \cite{IMJ}. We suggest another version of  the quantum billiard approach
 by   deducing asymptotic solutions to WDW equation which obey
 the master equation with anomaly term $A < 0$. We consider two examples of
 quantum billiards: triangle $2$-dimensional
 Chitre's billiard  for $D = 4$  model with three $2$-forms and
 $9$-dimensional billiard for $D = 11$ model with $120$ $4$-forms
 which mimics $M2$-brane sector of $D=11$ supergravity. For certain
 asymptotic solutions the vanishing of the wave function is proved.

  \section{The model}

 Here we consider  the multidimensional gravitational
 model governed by the  action
 \beq{2.1}
 S_{act} = \frac{1}{2\kappa^{2}}
 \int_{M} d^{D}z \sqrt{|g|} \{ {R}[g]
  - \sum_{s \in S} \frac{\theta_s}{n_s!} (F^s)^2 \}
  + S_{YGH},
 \eeq
where $g = g_{MN} dz^{M} \otimes dz^{N}$ is the metric on the
manifold $M$, ${\dim M} = D$,  $\theta_s  \neq 0$, $F^s =  dA^s
=\frac{1}{n_s!} F^s_{M_1 \ldots M_{n_a}}  dz^{M_1} \wedge \ldots
\wedge dz^{M_{n_s}}$ is a $n_s$-form ($n_s \geq 2$) on a
$D$-dimensional manifold $M$,  $s \in S$. In (\ref{2.1}) we denote
$|g| = |\det (g_{MN})|$,
 $(F^s)^2 =
  F^s_{M_1 \ldots M_{n_s}} F^s_{N_1 \ldots N_{n_a}}
  g^{M_1 N_1} \ldots g^{M_{n_s} N_{n_s}}$,
$s \in S$, where $S$ is some finite set of indices and $S_{\rm
YGH}$ is the standard York-Gibbons-Hawking boundary term
\cite{Y,GH}.

In the models with one time and usual form fields all $\theta_s >
0$ when the signature of the metric is $(-1,+1, \ldots, +1)$. For
this choice of the signature  $\theta_s < 0$ corresponds to  a
phantom form field $F^s$.

\subsection{Ansatz for non-composite brane configurations}

Let us consider the manifold
 \beq{2.10}
  M = \R_{*}  \times \R^{n},
 \eeq
with the metric
 \beq{2.11}
  g=  w e^{2{\gamma}(u)} du \otimes du   +
  \sum_{i=1}^{n} e^{2\phi^i(u)} \varepsilon(i) dx^i \otimes dx^i ,
 \eeq
where $\R_{*} = (u_{-}, u_{+})$, $w = \pm1$ and $\varepsilon(i) =
\pm 1$,  $i=1,\ldots,n$. The dimension of $M$ is $D = 1 + n$. For
$w = -1$ and $\varepsilon(i) = 1$, $i=1,\ldots,n$, we deal with
cosmological solutions while for $w = 1$,  and $\varepsilon(1) =
-1$, $\varepsilon(j) = 1$, $j= 2,\ldots,n$, we  get static
solutions (e.g. wormholes etc).

Let $\Omega = \Omega(n)$  be a set of all non-empty subsets of $\{
1, \ldots,n \}$. For any $I = \{ i_1, \ldots, i_k \} \in \Omega$,
$i_1 < \ldots < i_k$, we denote
  $\tau(I) \equiv dx^{i_1}  \wedge \ldots \wedge dx^{i_k}$,
  $\eps(I) \equiv \eps(i_1) \ldots \eps(i_k)$,
  $d(I) = |I| \equiv k$ .

For the fields of forms we consider the following non-composite
electric ansatz
 \beq{2.1.1}
  A^s= \Phi^s \tau(I_s), \qquad F^s= d\Phi^s \wedge\tau(I_s),
 \eeq
where  $\Phi^s=\Phi^s(u)$ is smooth function on $\R_{*}$ and $I_s
\in \Omega$, $s \in S$. Due to (\ref{2.1.1}) we have
$d(I_s)=n_s-1$, $s \in S$.

\subsection{Sigma-model action}

It was proved in \cite{IMC} that the equations of motion for the
model (\ref{2.1})  with the fields from (\ref{2.11}) and
(\ref{2.1.1}) are equivalent to equations of motion for the
$\sigma$-model governed by the action
 \beq{2.2.7}
  S_{\sigma 0} =  \frac{\mu}{2}
  \int du {\cal N}  \biggl\{ G_{ij}
  \dot \phi^i  \dot \phi^j
  + \sum_{s\in S} \eps_s \exp{(-2 U_i^s \phi^i)}
  (\dot \Phi^s)^2 \biggr\},
 \eeq
where $\dot X \equiv dX/du$,  $\mu \neq 0$, $\gamma_0(\phi) \equiv
\sum_{i=1}^n \phi^i$ and  ${\cal N}=\exp(\gamma_0-\gamma)>0$ is
modified lapse function,
  \beq{2.2.10}
   G = G_{ij} d \phi^i\otimes d\phi^j, \qquad   G_{ij}=  \delta_{ij} - 1,
  \eeq
is truncated target space metric and co-vectors $U^s$ read
 \beq{2.2.11}
   U^s(\phi) =  U_i^s \phi^i = \sum_{i \in I_s} \phi^i,
  \quad
  U^s = (U_i^s) =  \delta_{iI_s},
 \eeq
 $s \in S$.

 Here
 \beq{2.2.12}
  \delta_{iI}=\sum_{j\in I}\delta_{ij}
 \eeq
 is an indicator of $i$ belonging
 to $I$: $\delta_{iI}=1$ for $i\in I$ and $\delta_{iI}=0$ otherwise; and
 \beq{2.2.13a}
   \eps_s= \eps(I_s) \theta_{s},
 \eeq
  $s\in S$.

  In what follows we will use the scalar product
 \beq{3.1.1}
  (U,U')= G^{ij} U_i U'_j,
 \eeq
 for $U = (U_i), U' = (U'_i) \in \R^{n}$,  where
 $(G^{ij})$ is the matrix inverse to  the matrix $( G_{ij})$
  \beq{3.1.3}
    G^{ij}= \delta^{ij} +\frac1{2-D},
 \eeq
 $i,j=1,\dots,n$.

The scalar products of $U$-vectors  (\ref{2.2.11}) read \cite{IMC}
 \beq{3.1.4}
 (U^s,U^{s'})=d(I_s\cap I_{s'})+ \frac{d(I_s)d(I_{s'})}{2-D},
 \eeq
 $s,s' \in S$.

Action (\ref{2.2.7}) may be also written in the form
 \beq{4.1.6}
  S_\sigma=\frac{\mu}{2} \int du{\cal N}\left\{
  {\cal G}_{A B}(X)\dot X^{ A}\dot X^{ B} \right\},
  \eeq
where $X = (X^{ A})=(\phi^i,\Phi^s)\in {\R}^{N}$, $N = n + m$, $m
= |S|$ is the number of branes  and minisupermetric    ${\cal G}=
{\cal G}_{ A B}(X)dX^{ A}\otimes dX^{ B}$ on minisuperspace ${\cal
 M}={\bf R}^{N}$ is defined by the relation

  \beq{3.2.3g}
  {\cal G}= G + \sum_{s\in S}\eps_s
  \e^{-2U^s(\phi)} d\Phi^s\otimes d\Phi^s,
  \eeq
where $G$ is defined in (\ref{2.2.10}) and $U^s(\phi)=U_i^s
\phi^i$ is defined in (\ref{2.2.11}).

In what follows we use the notation
 \beq{4.1.11}
  U^{\Lambda}(\phi)=U_i^{\Lambda} \phi^i= \gamma_0(\phi),
  \qquad U^{\Lambda}_i = 1.
 \eeq
The  vector  $U^{\Lambda} = (U^{\Lambda}_i)$ is time-like, since
\cite{IMC}
 \beq{4.1.15}
 (U^{\Lambda},U^{\Lambda})=-\frac{D-1}{D-2} < 0.
 \eeq

\section{Quantum billiard approach}

In this section  we develop a quantum analogue of the billiard
approach which is understood as finding of certain asymptotic
solutions to Wheeler--DeWitt (WDW) equation.

\subsection{Restrictions.}

First we outline  two restrictions  which will be used in
derivation of the quantum billiard
  \bear{3.1.4d}
  (i) \quad d(I_s) < D -2, \\ \label{3.1.4e}
  (ii) \quad \quad \quad  \quad \eps_s > 0,
  \ear
 for all $s$. The first restriction on the dimensions of the brane
 worldvolumes  excludes domain walls.
 The second one is a necessary condition for the formation
 of infinite ``wall'' potential in certain limit (see below).

 Due to the first restriction we get
 \beq{3.1.4u}
  (U^s,U^{s})=d(I_s) \left(1 +\frac{d(I_{s})}{2-D} \right)  > 0,
  \quad s \in S.
 \eeq

\subsection{Wheeler-DeWitt equation}

 Let us fix the temporal gauge as follows
 \beq{4.2.1}
  \gamma_0-\gamma= 2 f(X),  \quad  {\cal N} = e^{2f},
 \eeq
where $f$: ${\cal M}\to{\bf R}$ is a smooth function. Then we obtain the
Lagrange system with the Lagrangian
 \beq{4.2.3}
   L_f=\frac\mu2\e^{2f}{\cal G}_{A B}(X)
   \dot X^{ A}\dot X^{ B}
 \eeq
and the energy constraint
 \beq{4.2.4}
  E_f=\frac{\mu}{2} \e^{2f}{\cal G}_{ A  B}(X)
  \dot X^{ A}\dot X^{ B}=0.
 \eeq

Using the standard prescriptions of (covariant and conformally
covariant)  quantization of the energy constraint, see \cite{Mis,
Hal,IMZ,IMJ} and refs. therein,  we are led to the Wheeler-DeWitt
(WDW) equation
 \beq{4.2.5}
  \hat{H}^f \Psi^f \equiv
  \left(-\frac{1}{2\mu}\Delta\left[e^{2f}{\cal G}\right]+
  \frac{a}{\mu} R\left[e^{2f}{\cal G}\right]
  \right)\Psi^f=0,
  \eeq
where
 \beq{4.2.5a}
  a=a_N= \frac{(N-2)}{8(N-1)},
 \eeq
  $N = n + m$.

Here $\Psi^f = \Psi^f(X)$ is the wave function corresponding to
the $f$-gauge (\ref{4.2.1}) and satisfying the relation
 \beq{4.2.7}
  \Psi^f= e^{bf} \Psi^{f=0}, \quad b = (2-N)/2.
 \eeq

 In (\ref{4.2.5}) we denote by $\Delta[{\cal G}^f]$ and
 $R[{\cal G}^f]$  the Laplace-Beltrami operator and the scalar
 curvature corresponding to the metric
 \beq{4.2.7G}
 {\cal G}^f =   e^{2f} {\cal G},
 \eeq
 respectively.

 Let us put $f = f(\phi)$. Then we get
 \bear{4.2.9}
  \tri[{\cal G}^f]
  =  \e^{\bar{U}}|\bar{G}|^{-1/2} \frac\partial{\partial \phi^i}\left(\bar{G}^{ij}
  \e^{- \bar{U}} |\bar{G}|^{1/2} \frac\partial{\partial \phi^j}\right)
   \\ \nonumber
  +  \sum_{s\in S} \e^{2 \bar{U}^s(\phi)}
  \left(\frac\partial{\partial\Phi^s}\right)^2,
  \ear
 where
   \beq{4.2.9U}
   \bar{U}= \sum_{s\in S} \bar{U}^{s}, \qquad  \bar{U}^{s} = U^{s}(\phi) - f
   \eeq
    and
 \beq{4.2.9G}
   \bar{G}_{ij} =  e^{2f} G_{ij}, \qquad \bar{G}^{ij} =  e^{-2f}
    G^{ij},
 \eeq
 $|\bar{G}| = |\det{(\bar{G}_{ij})}|$).

  Here we are interested in a special class of asymptotical solutions to
  WDW-equation. The metrics  $G$, $\cal{G}$ have  pseudo-Euclidean
   signatures $(-,+,...,+)$ (the last one - due to (\ref{3.1.4e})). We put
  \beq{5.1}
     e^{2f} = - (G_{ij}\phi^i \phi^j)^{-1},
  \eeq
   where $G_{ij}\phi^i \phi^j < 0$.

   In what follows we will use a diagonalization of
   $\phi$-variables
 \beq{5.1.z}
   \phi^{i} = S^{i}_{a}z^{a},
  \eeq
$a = 0, ..., n-1$,   obeying $G_{ij}\phi^i
 \phi^j  = \eta_{ab} z^{a}z^{b}$, where $(\eta_{ab}) =
{\diag}(-1,+1,...,+1)$.

 We restrict  the WDW equation to the lower light cone
   $V_{-} = \{z = (z^{0}, \vec{z}) | z^{0} < 0, \eta_{ab} z^{a}z^{b} < 0 \}$
 and  introduce  Misner-Chitre-like coordinates
 \bear{5.2z}
     z^{0} = - e^{-y^{0}}\frac{1 + \vec{y}^{2}}{1 -\vec{y}^{2}}, \\
     \label{5.2zz}
     \vec{z} = - 2 e^{-y^{0}} \frac{\vec{y}}{1 - \vec{y}^{2}},
  \ear
  where $y^{0} < 0$ and $\vec{y}^{2} < 1$.
  We note that in these variables $f = y^{0}$.

  Using the relation $f_{,i} =  \bar{G}_{ij} \phi^{j}$,
 following from   (\ref{5.1}), we obtain
  \beq{5.3}
       \tri[\bar{G}]f = 0, \qquad \bar{G}^{ij} f_{,i} f_{,j} = -1.
  \eeq
 These relations may readily deduced from
 the following formula
 \beq{5.4}
 \bar{G} = - dy^{0} \otimes d y^{0} + h_L,
 \eeq
   where
 \beq{5.5}
    h_L =  \frac{4 \delta_{rs} dy^{r} \otimes dy^{s}}{(1 - \vec{y}^{2})^2},
 \eeq
  (with summation over $r,s = 1,..., n -1$ assumed). Here the
  metric $h_L $ is defined on the unit ball
  $D^{n -1} = \{ \vec{y} \in \R^{n -1}| \vec{y}^{2} < 1
  \}$. The pair $(D^{n -1}, h_L)$ is one of the realization of
  $(n -1)$-dimensional analogue of the Lobachevsky space.

 For the wave function  we consider the following ansatz
      \beq{5.6}
         \Psi^f = e^{C(\phi)} \Psi_{*},
       \eeq
where  the pre-factor $e^{C(\phi)}$ is chosen in order to cancel
the terms linear in derivatives ($\Psi_{*,i}$) arising in
calculation of   $\tri[{\cal G}^f] \Psi^f$. This takes place if we
put
      \beq{5.7}
      C(\phi) = \frac{1}{2} \bar{U} =
      \frac{1}{2}(\sum_{s\in S} U^{s}_i \phi^i - m f).
      \eeq

     From (\ref{5.6}) and (\ref{5.7}) we get
     \bear{5.12}
    \left(-\frac{1}{2}\Delta\left[e^{2f}{\cal G}\right]+
    a  R\left[e^{2f}{\cal G}\right]
     \right) (e^{C(\phi)} \Psi_{*}) =
  \\ \nn
     e^{C(\phi)}
     \left(-\frac{1}{2}\Delta\left[\bar{G} \right] -
      \frac{1}{2} \sum_{s\in S} \e^{2 \bar{U}^s}
       \left(\frac\partial{\partial\Phi^s}\right)^2 + \delta V \right)
       \Psi_{*},
  \ear
  where
     \beq{5.13}
      \delta V = A e^{-2f} - \frac{1}{8} (n-2)^2.
      \eeq
  Here we denote
          \beq{5.14}
      A  =   \frac{1}{8(N-1)} [ \sum_{s, s' \in
     S} (U^s,U^{s'}) - (N - 2) \sum_{s \in S} (U^s,U^{s}) ].
      \eeq

     Now we put
      \beq{5.15}
        \Psi^f = e^{C(\phi)} e^{iQ_s \Phi^s} \Psi_{0,L},
      \eeq

  where parameters $Q_s \neq 0$  correspond  to charge densities
  of branes and $e^{iQ_s \Phi^s} = \exp(i \sum_{s \in S} Q_s \Phi^s)$.
  Using relation (\ref{5.12}) we get

  \bear{5.16}
  \hat{H}^f \Psi^f = \mu^{-1} e^{C(\phi)}
  e^{iQ_s \Phi^s}  \left(-\frac{1}{2} \tri[\bar{G}]+  \right.\\
  \nn \qquad \qquad
    \left.  \frac{1}{2} \sum_{s \in S} Q_s^2 e^{-2f + 2U^s(\phi)})
    + \delta V \right) \Psi_{0,L}=0.
  \ear

\subsection{Asymptotic behavior of the solutions for $y^0 \to - \infty$
}

Here we deal with  asymptotic solutions to WDW equation in the
limit $y^0 \to - \infty$. Due to relations  (\ref{5.15}) and
(\ref{5.16}) this equation reads
  \beq{5.17}
    \left(-\frac{1}{2} \tri[\bar{G}]+
    \frac{1}{2} \sum_{s \in S} Q_s^2 e^{-2f + 2U^s(\phi)}
    + \delta V  \right) \Psi_{0,L}=0.
  \eeq

It was shown in our  paper on the classical billiard approach
\cite{IMb1} that
  \beq{5.18}
    \frac{1}{2} \sum_{s \in S} Q_s^2 e^{-2f + 2U^s(\phi)}
    \to V_{\infty},
  \eeq
as $y^0 = f  \to - \infty$.

   In this relation $V_{\infty}$ is the potential of infinite walls which
   are produced by branes:
      \beq{5.18a}
     V_{\infty} = \sum_{s \in S} \theta_{\infty}( \vec{v}_s^2 -1 - (\vec{y} - \vec{v}_s)^2 )
      \eeq

  Here we use the notation $\theta_{\infty}(x) = + \infty $ for $x \geq
  0$ and $\theta_{\infty}(x) = 0$ for $x < 0$. The vectors
  $\vec{v}_s$, $s \in S$, belonging to $\R^{n -1}$
  are defined by the formulae
      \beq{5.20}
      \vec{v}_s =  -  \vec{u}_s/u_{s0},
      \eeq
  where $n$-dimensional vectors  $u_s = (u_{s0},\vec{u}_s) = (u_{sa})$
  are obtained from $U^s$-vectors using a diagonalization matrix $(S^{i}_{a})$
  from  (\ref{5.1.z})
  \beq{5.21}
    u_{sa} = S^{i}_{a} U^s_i.
  \eeq
    Due to condition (\ref{3.1.4d})
  \beq{5.21a}
  (U^s,U^s) = -(u_{s0})^2 + (\vec{u}_s)^2 > 0
  \eeq
  for all $s$. Here
  we use a diagonalization (\ref{5.1.z}) from
  \cite{IMb1} obeying
  \beq{5.21b}
   u_{s0} > 0
  \eeq
  for all $s \in S$. The inverse matrix   $(S_{i}^{a}) =
  (S^{i}_{a})^{-1}$ defines the    the map inverse to (\ref{5.1.z})
   \beq{5.21c}
   z^{a} = S_{i}^{a} \phi^{i},
   \eeq
  $a = 0, ..., n -1$.

  The inequalities (\ref{5.21a})
 imply  $|\vec{v}_s| > 1$ for all $s$. The potential $V_{\infty}$
 corresponds to the billiard $B$ in the multidimensional Lobachevsky
 space $(D^{n -1}, h_L)$. This  billiard is an open domain in
 $D^{n -1}$ which is defined by a set of inequalities:
     \beq{5.22}
       |\vec{y} - \vec{v}_s| < \sqrt{\vec{v}_s^2 -1} = r_s,
     \eeq
  $s \in S$. The boundary $\partial B$ is formed by  parts of
  hyper-spheres with  centers in $\vec{v}_s$ and radii $r_s$.

 The condition (\ref{5.21b}) is also obeyed for the
 diagonalization (\ref{5.21c}) with
 \beq{5.21cc}
 z^{0} = U_i \phi^{i}/\sqrt{|(U,U)|},
 \eeq
 where $U$-vector is time-like
 \beq{5.21U}
 (U,U) < 0
 \eeq
 and
 \beq{5.21Us}
 (U,U^s) < 0
 \eeq
 for all $s \in S$.

 The inequalities (\ref{5.21U}) and (\ref{5.21Us})
 are satisfied identically if $U = k U^{\Lambda}$,  $k > 0$,
 see  (\ref{4.1.11}).

 Conditions (\ref{5.21Us}) and   hence   (\ref{5.21b}) may be
 relaxed.  In this case we obtain a more general prescription for
 the drawing of the billiard walls (e.g. for $u_{s0} < 0$  and
 $u_{s0} = 0$) described in \cite{IMb-09}.

 Thus, we are led to an asymptotic relation for the function
 $\Psi_{0,L}(y^0,\vec{y})$
 \beq{5.23}
 \left(-\frac{1}{2}\tri[\bar{G}]+ \delta V \right) \Psi_{0,L}=0
 \eeq
 with $\vec{y} \in B$ and the zero boundary condition $\Psi_{0,L|\p B} = 0$
imposed. Due to (\ref{5.4}) we get   $\tri[\bar{G}] = - (\p_0)^2 +
 \tri[h_L]$, where $\tri[h_L] = \Delta_{L}$ is the Laplace-Beltrami operator
 corresponding to the  $(n-1)$-dimensional  Lobachevsky metric $h_L$.

 By splitting the variables
  \beq{5.24}
   \Psi_{0,L}= \Psi_{0}(y^0) \Psi_{L}(\vec{y})
   \eeq
 we are led to the asymptotic relation (for $y^{0} \rightarrow -
   \infty$)
 \beq{5.25}
 \left( \left(\frac{\partial}{\partial y^{0}}\right)^{2} - \Delta_{L} +
 2Ae^{-2y^{0}} +  E - \frac{1}{4} (n -2)^2 \right)\Psi_{0} = 0
\eeq
 equipped with the relations

 \beq{5.26}
 \Delta_{L}\Psi_{L} = - E \Psi_{L}, \qquad  \Psi_{L| \p  B}=0.
 \eeq

 Here we assume that the  operator $(-\Delta_{L})$  with the zero
 boundary condition imposed has a spectrum obeying
   \beq{5.27}
   E \geq  \frac{1}{4} (n -2)^2.
   \eeq
This is valid at least when the  billiard $B$ is (sub)compact and
small  enough. The examples of non-(sub)compact billiards obeying
(\ref{5.27}) are considered in the next section.

 Here we  put
 \beq{5.27a}
  A < 0.
 \eeq
 Solving equation (\ref{5.25}) we get for $A < 0$   the following basis of solutions
 \beq{5.28}
 \Psi_{0} = {\cal B}_{i \omega} \left(\sqrt{2|A|}e^{-y^{0}}\right),
  \eeq
 where  ${\cal B}_{i \omega}(z) = I_{i \omega}(z), K_{i\omega}(z)$ are modified Bessel
 functions and
 \beq{5.29}
 \omega = \sqrt{E -  \frac{1}{4} (n -2)^2} \geq 0.
 \eeq

 In semi-quantum case  (with quantum gravity and  classical matter source)
 \cite{IMb0,KKN,KN} the anomaly term is absent, i.e. $A=0$.

 Using asymptotical relations
 \beq{5.30}
   I_{\nu} \sim \frac{e^z}{\sqrt{2 \pi z}}, \qquad  K_{\nu} \sim \frac{e^{-z}}{\sqrt{2 z}},
   \eeq
  for $z \to + \infty$, we get
  \beq{5.31}
 \Psi_{0} \sim C_{\pm} \exp \left( \pm \sqrt{2|A|}e^{-y^{0}} + \frac{1}{2} y^{0} \right),
 \eeq
 for $y^0 \to - \infty$. Here  $C_{\pm}$ are nonzero  constants and  ``+''
 corresponds to ${\cal B} = I$ and ``-'' - to ${\cal B} = K$.
 Now we evaluate the pre-factor $e^{C(\phi)}$ in (\ref{5.15}), where
 \beq{5.32}
 C(\phi) = \frac{1}{2}(U(\phi) - m f)
 \eeq
 and
 \beq{5.33}
 U(\phi) = U_i \phi^i =  \sum_{s\in S} U^{s}_i
 \phi^i, \qquad U_i  =  \sum_{s\in S} U^{s}_i.
 \eeq

 Now we use $U = (U_i )$ as a time-like vector which defines  $z^0$  in
 (\ref{5.21cc}).  Thus, we need to impose the restriction $(U,U) < 0$.
Then, using (\ref{5.21cc}), (\ref{5.2z})  and $f = y^0$  we obtain
 \beq{5.34}
  C(\phi) =   \frac{1}{2} \left( - q e^{-y^{0}}
  \frac{(1 + \vec{y}^{2})}{1 -\vec{y}^{2}}  - m y^0
 \right),
\eeq
where
 \beq{5.34q}
 q = \sqrt{-(U,U)} > 0.
 \eeq

Combining relations (\ref{5.15}), (\ref{5.24}), (\ref{5.31}) and
(\ref{5.34}) we find
 \beq{5.35}
   \Psi^f
  \sim C_{\pm} \exp \left( \theta^{\pm}(|\vec{y}|)e^{-y^{0}}
   - \frac{1}{2}(m -1) y^{0}
  \right) e^{iQ_s \Phi^s} \Psi_{L}(\vec{y}),
 \eeq
 as $y^0 \to - \infty$  for fixed $\vec{y} \in B$.
 Here  $C_{\pm} \neq 0$ and
\beq{5.36}
 \theta^{\pm}(|\vec{y}|) = - \frac{q}{2}   \frac{(1 + \vec{y}^{2})}{(1 -\vec{y}^{2})}
 \pm \sqrt{-2A},
\eeq
 where ``+'' corresponds to the Bessel function ${\cal B} = I$ and ``-'' -
to ${\cal B} = K$.

Relation (\ref{5.14}) may be rewritten as
 \beq{5.14A}
 A  =   \frac{1}{8(N-1)} [ (U,U) - (N - 2) \sum_{s \in S} (U^s,U^{s}) ].
 \eeq
 where we have used  identity
 \beq{5.37}
 (U,U) = \sum_{s, s' \in  S} (U^s,U^{s'})
  \eeq
 following from the definition of $U$ in   (\ref{5.33}). It should be noted that
 restrictions $(U,U) < 0$ and $(U^s,U^{s}) > 0$, $s \in S$, imply $A
  < 0$.

 Now we study the asymptotical behaviour of the wave function  (\ref{5.15})

 \beq{5.37Psi}
 \Psi^f = e^{C(\phi)} e^{iQ_s \Phi^s} {\cal B}_{i \omega}
 \left(\sqrt{2|A|}e^{-y^{0}}\right) \Psi_{L}(\vec{y}),
 \eeq
 with  $C(\phi)$ from (\ref{5.34}) and $(U,U) <  0$, $A < 0$.

 Let  i) ${\cal B} = K$. Then
 \beq{5.37P}
 \Psi^f  \to 0
 \eeq
as $y^0 \to - \infty$  for fixed $\vec{y} \in B$ and $\Phi^s \in
\R$, $s \in S$.  This  follows just from  (\ref{5.35}).

Now we consider the case  ii) ${\cal    B} = I$. First we put
\beq{5.38A}
  \quad  \frac{1}{2} q > \sqrt{2|A|},
 \eeq
We get $$\Psi^f   \to 0 $$ as $y^0 \to - \infty$ for fixed
$\vec{y} \in B$   and $\Phi^s \in \R$, $s \in  S$. This also
follows from (\ref{5.35}).

Let us consider the second case
 \beq{5.38Aa}
 \quad \frac{1}{2} q = \sqrt{2|A|}.
\eeq
We obtain
$$\Psi^f  \to 0 $$
as $y^0 \to - \infty$
for fixed $\vec{y}  \in B \setminus \{ \vec{0} \} $ and $\Phi^s
 \in \R$, $s \in  S$.  This  also follows from (\ref{5.35}).
Moreover, $|\Psi^f| \to + \infty$ as $y^0 \to - \infty$,
when $\vec{y} = \vec{0}$ and $\Psi_{L}(\vec{0}) \neq 0$.

Now we consider the third case
\beq{5.38Ab} \frac{1}{2} q <
\sqrt{2|A|}.
\eeq
If the point $\{ \vec{0} \}$ belongs to the
billiard $B$ and $\Psi_{L}(\vec{0}) \neq 0$ then, it may be
readily verified that there exists $\eps >0$ such that for all
$\vec{y}$ obeying $|\vec{y}| < \eps$ (and all $\Phi^s \in \R$, $s
\in  S$)
 \beq{5.39}
 |\Psi^f | \to + \infty
 \eeq
as $y^0 \to - \infty$.

\section{Examples}

Here we consider two examples of quantum billiards in dimensions
$D = 11$ and $D = 4$. In this section we deal with $(n+
1)$-dimensional cosmological metrics  (\ref{2.11}) with $w = -1$.

\subsection{$9$-dimensional billiard in $D=11$ model}

Let us consider $11$-dimensional gravitational model with several
4-forms, which gives non-composite analogous of $M$-brane
solutions in $D = 11$ supergravity \cite{CJS}. The action  reads
as follows
 \beq{6.1}
  S_{11} = \frac{1}{2\kappa^2_{11}}
  \int_{M} d^{11}z \sqrt{|g|} \{ {R}[g] + {\cal L}  \}+ S_{YGH},
 \eeq

where
 \beq{6.2}
 {\cal L} =
  -  \frac{1}{4!} \sum_{I \in S}  (F^I_{4})^2.
 \eeq

Here $F^I_{4}$ is  4-form with the index $I \in S$, where here $S
$ is the set of all subsets with three elements:   $I = \{ i_1,
i_2, i_3 \}$, $1 \leq i_1 < i_2 < i_3 \leq 10$. The number of
elements in $S$ is $120$.

The action (\ref{6.1}) with  $\cal{L}$ from (\ref{6.2}) mimics
non-composite $SM2$-brane solutions which are given by the
 metric  (\ref{2.11}) with $w = -1$, $n=10$ and
 \beq{6.2e}
 F^I_{4} =   d\Phi^I \wedge\tau(I),
 \eeq
$I \in S$.

We consider the non-trivial case when all charge densities of
branes $Q_I$, $I \in S$, are non-zero. In  the classical case we
get  a $9$-dimensional billiard $B \in H^9$ with $120$
``electric'' walls. This billiard was found in \cite{DamH1}. It
has a finite volume.

The minus Laplace-Beltrami   operator $(-\Delta_{L})$ on $B$ with
zero boundary conditions has a  spectrum obeying restriction
(\ref{5.27}) with $n = 10$ \cite{KN}.

Let us us calculate $(U,U)$, where $U = \sum_{s \in S}
 U^s$. We get $U_i = \sum_{I \in S} \delta_{iI}$ where
 $\delta_{iI}$  is defined in (\ref{2.2.12}). Thus, $U_i$ is the
number of sets $I \in S$ which contain the point $i$. It is
obvious  that $U_i = C_9^2 = 36$. Thus, $U = 36 U^{\Lambda}$ and
hence we may use the $z$-variables from \cite{{IMb0},{IMb1}}.

 We get (see (\ref{3.1.3}))
 \beq{6.3}
 (U,U) = G^{ij}U_i U_j =  \sum_{i,j = 1}^{10} (\delta^{i j} -
 \frac{1}{9})(36)^2= - 1440 < 0
 \eeq
in agreement with our restriction (\ref{5.21U}). Since $N = 130$
($m=120$) and all $(U^s,U^s) =2$ we obtain from  (\ref{5.14A}) the
following value for the anomaly number
 \beq{6.3A}
 A =  - \frac{1340}{43}.
 \eeq

 In this case the inequality (\ref{5.38A}) is satisfied and hence
 we get from the previous analysis that
 the wave function $\Psi^f \to 0$ is asymptotically vanishing
 as $y^0 \to - \infty $.

\subsection{$2$-dimensional billiard in $D=4$ model}

Let us consider the $4$-dimensional gravitational model with three
$2$-forms which gives the two-dimensional Chitre's billiard.

The action  reads
 \beq{7.1}
  S_{4} = \frac{1}{2\kappa^2}
  \int_{M} d^{4}z \sqrt{|g|} \{ {R}[g] + {\cal L}  \}+ S_{YGH},
 \eeq

where
 \beq{7.2}
  {\cal L}_e =
   -  \frac{1}{2} \sum_{i = 1,2,3 }  (F^i_{2})^2.
 \eeq

Here $F^i_{2}$ is a 2-form, $i =1,2,3$.

 By using the ansatz
 \beq{7.3}
 F^i_{2} = d \Phi^i(u) \wedge dx^i,
 \eeq
 $i =1,2,3$, with non-zero charges $Q_i$ and the
 metric  (\ref{2.11}) with $w = -1$, $n=3$,
 we are led to the Chitre's triangle billiard (see \cite{IMb1}) which has a finite
 volume . The energy restriction (\ref{5.27}) is also obeyed in this
 case \cite{KN}.

 The calculations give us $(U^s,U^{s}) = \frac{1}{2}$,  $(U,U) = - \frac{3}{2}$ and
 \beq{7.4}
 \qquad A = - \frac{3}{16}.
 \eeq

 According to the analysis which was performed above we get
 the asymptotic vanishing of  the wave function:
 $\Psi^f \to 0$   as $y^0 \to - \infty$, when either
 i) ${\cal B} = K$, or ii) ${\cal B} = I$ and
 $\vec{y} \neq \vec{0}$ (see (\ref{5.38Aa})).

\section{Conclusions}

Here we have considered the quantum billiard approach for the
cosmological-type model with $n$ one-dimensional factor-spaces in
the theory with several forms. When electric non-composite brane
ansatz was adopted and certain restrictions on parameters of the
model were imposed the Wheeler-DeWitt (WDW) equation for the
model, written in conformally-covariant form, was analyzed.

We have imposed  certain restrictions on parameters of the model
and have obtained asymptotic solutions to WDW equation. These
solutions are of quantum billiard  form since they are governed by
the spectrum of the Lapalace-Beltrami operator on the billiard
with the zero boundary condition imposed. The billiard belongs to
the $(n-1)$-dimensional Lobachevsky space $H^{n-1}$.

We have presented two examples of quantum billiards : (a) the
quantum $d=2$ billiard  in $D=4$ gravitational model with three
$2$-forms and (b) the quantum $d=9$ billiard for $D = 11$
gravitational model with $120$ $4$-forms which mimics the quantum
billiard with $M2$-branes in $D=11$ supergravity. We have shown
 the asymptotic vanishing of the wave function:  $\Psi^f  \to 0$,
  in the case (b)  for all basis solutions and in the case (a) for the
 Bessel function ${\cal B} = K$ and for  ${\cal B} = I$ when
 $\vec{y} \neq \vec{0}$.



\small

\begin{thebibliography}{99}

\bibitem{Chit}
 D.M. Chitre, Ph. D. Thesis (University of Maryland) 1972.

 \bibitem{BLK}
 V.A. Belinskii, E.M. Lifshitz  and I.M. Khalatnikov, {\it Usp.
 Fiz. Nauk}  {\bf 102}, 463 (1970) [in Russian]; {\it Adv. Phys.}
 {\bf 31}, 639 (1982).

 \bibitem{Mis0}
  C.W. Misner,  Quantum cosmology, {\it Phys. Rev.} {\bf 186}, 1319 (1969).

 \bibitem{Mis1}
 C.W. Misner,  The Mixmaster cosmological metrics, preprint UMCP
 PP94-162; gr-qc/9405068.

  \bibitem{Kir-93}
 A.A. Kirillov, {\it Sov. Phys.} {\it JETP} {\bf 76}, 355 (1993)
 [ {\it ZhETF} {\bf 76}, 705 (1993), in Russian];
 {\it Int. Jour. Mod. Phys.} {\bf D3}, 431 (1994).

 \bibitem{IKM1}
 V.D. Ivashchuk, A.A. Kirillov and V.N. Melnikov,
 On Stochastic   Properties of Multidimensional Cosmological Models near the
 Singular Point,   {\it Izv. Vuzov (Fizika)} {\bf 11}, 107 (1994) (in Russian)
 [Russian Physics Journal  {\bf 37}, 1102 (1994)].

 \bibitem{IKM2}
 V.D. Ivashchuk, A.A. Kirillov and V.N. Melnikov,
 On Stochastic   Behaviour of Multidimensional Cosmological Models near the
 Singularity,   {\it Pis'ma ZhETF } {\bf 60},  No 4, 225 (1994) (in Russian)
 [JETP Lett. {\bf 60}, 235 (1994)].

 \bibitem{IMb0}
 V.D. Ivashchuk and V.N. Melnikov, Billiard Representation for
 Multidimensional Cosmology with Multicomponent Perfect Fluid near
 the Singularity,  {\it Class. Quantum Grav.} {\bf 12}, No 3,
 809-826 (1995); gr-qc/9407028.

 \bibitem{IMb1}
 V.D. Ivashchuk and V.N. Melnikov, Billiard Representation for
 Multidimensional Cosmology with Intersecting p-branes near the
 Singularity, {\it J. Math. Phys.}  {\bf 41}, No 9, 6341-6363 (2000);
 hep-th/9904077.

 \bibitem{DamH1}
 T. Damour and  M. Henneaux,
 Chaos in Superstring Cosmology,
 {\it Phys. Rev. Lett.} {\bf 85}, 920-923 (2000);
 hep-th/0003139.

 \bibitem{DHN}
 T. Damour, M. Henneaux and H. Nicolai,
 Cosmological billiards, topical review, {\it Class. Quantum
 Grav.} {\bf 20}, R145-R200 (2003); hep-th/0212256.

 \bibitem{KKN}
  A. Kleinschmidt, M. Koehn and  H. Nicolai,
  Supersymmetric quantum cosmological billiards,
  {\it Phys. Rev.} {\bf D 80}: 061701 (2009);
   arxiv: 0907.3048.

  \bibitem{KN}
  A. Kleinschmidt and H. Nicolai, Cosmological quantum billiards,
  arxiv: 0912.0854.

 \bibitem{LMMP}
 H. L\"u, J. Maharana, S. Mukherji  and C.N. Pope, Cosmological
 Solutions, p-branes and the Wheeler De Witt Equation,
 {\it Phys. Rev.} {\bf D 57}, 2219-2229 (1997); hep-th/9707182.

 \bibitem{IMJ}
 V.D. Ivashchuk and V.N. Melnikov,
 Multidimensional classical and quantum cosmology
 with intersecting $p$-branes, {\it J. Math. Phys.}
 {\bf 39}, 2866-2889 (1998); hep-th/9708157.

 \bibitem{Y}
 J.W. York, Role of conformal three-geometry in the dynamics of
 gravitation, {\it Phys. Rev. Lett.} {\bf 28} (16), 1082 (1972).

 \bibitem{GH}
 G.W. Gibbons and S.W. Hawking, Action integrals and partition
 functions in quantum gravity, {\it Phys. Rev.} {\bf D 15}, 2752
 (1977).

 \bibitem{IMC}
 V.D. Ivashchuk and V.N. Melnikov, Sigma-model for the
 Generalized  Composite p-branes,
 {\it Class. Quantum Grav.} {\bf 14}, 3001-3029 (1997);
 Corrigendum  {\it ibid.} {\bf 15} (1998) 3941-3942; hep-th/9705036.

 \bibitem{Mis}
 C.W. Misner, In: "Magic without Magic: John Archibald Wheeler", ed.
 J.R. Klauder, Freeman, San Francisko, 1972.

 \bibitem{Hal}
 J.J. Halliwell,
 Derivation of the Wheeler-De Witt Equation from a Path Integral
 for Minisuperspace Models, {\it Phys. Rev.} {\bf D 38}, 2468 (1988).

 \bibitem{IMZ}
 V.D. Ivashchuk, V.N. Melnikov and  A.I. Zhuk, On Wheeler-DeWitt
 Equation in Multidimensional Cosmology, {\it Nuovo  Cimento},
 {\bf B 104}, No 5, 575-581 (1989).

 \bibitem{IMb-09}
 V.D. Ivashchuk and  V.N. Melnikov,
 On billiard approach in multidimensional cosmological models,
 {\it Grav.  Cosmol.} {\bf 15}, No. 1, 49-58 (2009); arXiv: 0811.2786.

\bibitem{CJS}
 E. Cremmer, B. Julia and J. Scherk,
 Supergravity Theory in Eleven-Dimensions,
 {\it Phys. Lett.} {\bf B 76}, 409-412 (1978).

\end{thebibliography}
\end{document}